\documentstyle[epsf]{aipproc}

\let\BadCite=\cite
\def\cite{~\BadCite}
\catcode`@=11

\def\ifundefined#1{\expandafter\ifx\csname#1\endcsname\relax}
\def\citenum#1{\ifundefined{b@#1}{\bf#1}%
   \immediate\write16{citenum: Undefined argument #1}%
   \else\csname b@#1\endcsname\fi}

\catcode`@=12
\def\dofig#1#2{\epsfxsize=#1\centerline{\epsfbox{#2}}}

\def\sgn{\mathop{\rm sgn}}

\def\fbi{{\rm fb}^{-1}}

\def\lsp{{\tilde\chi_1^0}}
\def\tchi{{\tilde\chi}}

\def\mhalf{m_{1/2}}

\def\GeV{{\rm GeV}}

\def\Fig#1{Figure~\ref{#1}}

\def\slashchar#1{\setbox0=\hbox{$#1$}           % set a box for #1
   \dimen0=\wd0                                 % and get its size
   \setbox1=\hbox{/} \dimen1=\wd1               % get size of /
   \ifdim\dimen0>\dimen1                        % #1 is bigger
      \rlap{\hbox to \dimen0{\hfil/\hfil}}      % so center / in box
      #1                                        % and print #1
   \else                                        % / is bigger
      \rlap{\hbox to \dimen1{\hfil$#1$\hfil}}   % so center #1
      /                                         % and print /
   \fi}                                         %

\def\simge{%  ``greater than about'' symbol
    \mathrel{\rlap{\raise 0.511ex
        \hbox{$>$}}{\lower 0.511ex \hbox{$\sim$}}}}
\def\simle{%  ``less than about'' symbol
    \mathrel{\rlap{\raise 0.511ex
        \hbox{$<$}}{\lower 0.511ex \hbox{$\sim$}}}}

\font\twelvess=cmss10 scaled \magstep1

\begin{document}

\begingroup
\parindent=20pt
\thispagestyle{empty}
\vbox to 0pt{
\vskip-.25in
\moveleft.25in\vbox to 8.9in{\hsize=6.5in
{
\centerline{\twelvess BROOKHAVEN NATIONAL LABORATORY}
\vskip6pt
\hrule
\vskip1pt
\hrule
\vskip4pt
\hbox to \hsize{January, 1997 \hfil BNL-HET-98/6}
\vskip3pt
\hrule
\vskip1pt
\hrule
\vskip3pt

\vskip1in
\centerline{\LARGE\bf Sleptons at a First Muon Collider}
\vskip.5in
\centerline{\bf Frank E. Paige}
\vskip4pt
\centerline{Physics Department}
\centerline{Brookhaven National Laboratory}
\centerline{Upton, NY 11973 USA}

\vskip.75in

\centerline{ABSTRACT}
\vskip8pt
\narrower\narrower
	Signatures for sleptons, which have been extensively studied for
the Next Linear Collider, are reexamined taking into account some of the
different features of a First Muon Collider.

\vskip1in

	To appear in {\sl Workshop on Physics at the First Muon
Collider and at the Front End of a Muon Collider}, (Fermilab, November
6 -- 9, 1997).

\vskip0pt
}
\vfil\footnotesize
	This manuscript has been authored under contract number
DE-AC02-76CH00016 with the U.S. Department of Energy.  Accordingly,
the U.S.  Government retains a non-exclusive, royalty-free license to
publish or reproduce the published form of this contribution, or allow
others to do so, for U.S. Government purposes.
}
\vss}
\newpage

\endgroup
\setcounter{page}{1}

\title{Sleptons at a First Muon Collider}

\author{Frank E. Paige}
\address{Physics Department\\
Brookhaven National Laboratory\\
Upton, NY 11973}

\maketitle

\begin{abstract}
	Signatures for sleptons, which have been extensively studied for
the Next Linear Collider, are reexamined taking into account some of the
different features of a First Muon Collider.
\end{abstract}

	Supersymmetry (SUSY) signatures have been extensively studied
for the Next Linear Collider (NLC)\cite{NLC}. The basic
strategy\cite{TFMYO}\cite{NFT} uses the fact that SUSY particles are
produced in pairs and decay into Standard Model (SM) particles plus an
invisible lightest SUSY particle $\lsp$.  Hence the maximum and
minimum energies of the visible SM particles determine the initial
SUSY and LSP mass. However, the NLC studies have used properties of
the NLC such as easily variable energy and high electron
polarization.  This study makes assumptions appropriate for a First
Muon Collider (FMC), namely operation at a single energy with a
$20^\circ$ hole for shielding and no polarization. It does not,
however, take into account backgrounds from muon decays.

\section{Sleptons at LHC Point 5}

	This analysis is carried out for LHC Point 5, a minimal
supergravity (SUGRA) point with $m_0=100\,\GeV$, $\mhalf=300\,\GeV$,
$A_0=300\,\GeV$, $\tan\beta=2.1$, and $\sgn\mu=+1$. For this point,
$M(\lsp)=121.66\,\GeV$, $M(\tilde\ell_R)=157.20\,\GeV$,
$M(\tilde\ell_L)=238.82\,\GeV$, and $M(\tchi1\pm)=232.05\,\GeV$. The
LHC can trivially discover SUSY at this point and can use precision
measurements of combination of masses to determine the SUGRA
parameters. Using only such precision measurements, the estimated
errors for $10\,\fbi$ of luminosity are\cite{HPSSY}
\begin{itemize}
\item	$m_0 = 100.5{+12\atop-5}\,\GeV$,
\item	$\mhalf=298{+16\atop-9}\,\GeV$,
\item	$\tan\beta=1.8{+0.3\atop-0.5}$,
\item	$\sgn\mu=+1$.
\end{itemize}
$A_0$ is poorly determined because the weak-scale phenomenology is
insensitive to it. The ultimate LHC precision is considerably
better\cite{FROID}. Given these results, one would presumably choose
the energy of the FMC to be $\sqrt{s}=600\,\GeV$, the value assumed
here.

	SUSY and Standard Model events were generated as $e^+e^-$
events with ISAJET~7.31\cite{ISAJET}; $e$'s and $\mu$'s were
interchanged in the analysis. The toy detector simulation is the same
as that used for the LHC studies\cite{HPSSY} except that the $\eta$
range is limited to $|\eta|<1.8$, approximately equivalent to $\theta
> 20^\circ$. Jets were found with a fixed cone algorithm, and leptons
were taken from the generator. Slepton candidates are selected by
requiring two $\mu$ or $e$ leptons with $|\eta|<1.3$ and
$E_\ell>10\,\GeV$ and no other leptons or jets.  The two leptons are
required to satisfy
\begin{itemize}
\item	$E_\ell>10\,\GeV$, $|\eta_\ell|<1.3$ to select two identified
leptons in the detector,
\item	$|\vec p_1 + \vec p_2|< 0.9\sqrt{s}$ to reject lepton pair
background,
\item	$|\vec p_{T,1}+\vec p_{T,2}| > 10\,\GeV$ to reject lepton pair
and $\gamma\gamma$ background,
\item	$\Delta\phi_{1,2}<0.95\pi$ to reject lepton pair and
$\gamma\gamma$ background.
\end{itemize}
These cuts eliminate the $\ell^+\ell^-$, $\gamma\gamma \to
\ell^+\ell^-$, and $ZZ \to \ell^+\ell^-\nu\bar\nu$ backgrounds. A $Z$
mass cut was found to distort the $E_\ell$ distributions and was
replaced by the cut on $|\vec p_{T,1}+\vec p_{T,2}|$.

\begin{figure}[t]
\dofig{3in}{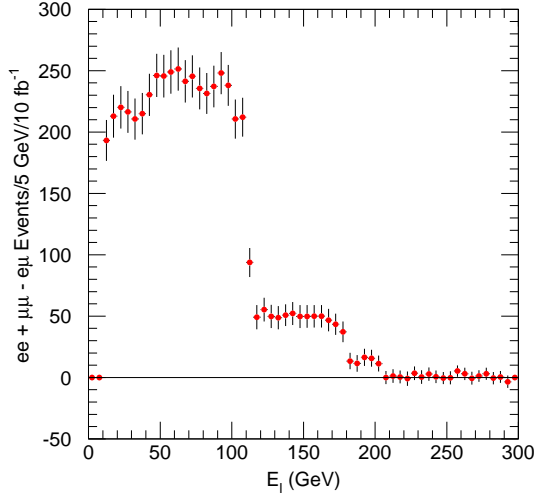}
\vskip6pt
\caption{The sum of the signal and background for $\mu^+\mu^- + e^+e^-
- \mu^\pm e^\mp$ at Point~5 with statistical errors appropriate for
$10\,\fbi$.\label{eleperr}}
\end{figure}

	After these cuts, the dominant background comes from leptonic
decays of $WW$ pairs. Since $W$'s decay equally into $e\nu$ and
$\mu\nu$, the SM background vanishes up to statistical fluctuations in
the combinations $\mu^+\mu^- + e^+e^- - e^+\mu^- - e^-\mu^+$ (and
also for $\mu^+\mu^- -e^+e^-$). This distribution is shown in
\Fig{eleperr} with error bars
appropriate for $10\,\fbi$ but with larger Monte Carlo statistics.

\begin{figure}[t]
\dofig{3in}{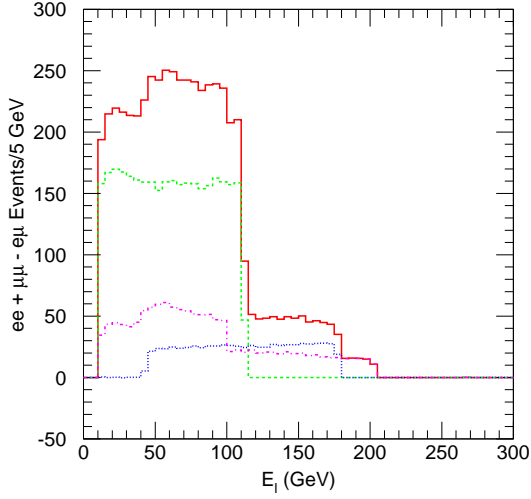}
\caption{Composition of $E_\ell$ distribution for $\mu^+\mu^- + e^+e^-
- e^+\mu^- - e^-\mu^+$ at Point~5. The dashed curve is from
$\tilde\ell_R\tilde\ell_R$; the dotted curve is from
$\tilde\ell_L\tilde\ell_L$; and the dashed-dotted curve is from
$\tilde\ell_R\tilde\ell_L$.\label{elepcomp}}
\end{figure}

	The origins of the signal are shown in \Fig{elepcomp}.  There
are contributions from $\tilde\ell_R\tilde\ell_R$,
$\tilde\ell_L\tilde\ell_L$, and $\tilde\mu+_R\tilde\mu_L$, the last
coming from gaugino exchange in the $t$-channel. Two-body kinematics
implies that for production of $\tilde\ell_i\tilde\ell_j$, $i,j=R,L$,
the maximum and minimum energies are
$$
E_{i\to\ell}^\pm = {M_i^2 - M_\lsp^2 \over 4 M_i^2} \left[{s+M_i^2-M_j^2
\pm \sqrt{(s-M_I^2-M_j^2)^2-4M_i^2M_j^2} \over 2\sqrt{s}}\right]
$$
There are four distinct ranges for the lepton energy, one each for
$RR$ and $LL$ events and two for $LR$ events:
\begin{eqnarray*}
E_{RR}	&=& (111.4\,\GeV, 11.2\,\GeV)\cr
E_{LL} 	&=& (178.3\,\GeV, 33.0\,\GeV)\cr
E_{LR}	&=& (99.5\,\GeV, 10.0\,\GeV)\cr
	&=& (101.9\,\GeV, 19.2\,\GeV)\cr
\end{eqnarray*}
If no other cuts had been made, one would obtain a sum of four square
distributions with these limits. All of these limits can be seen as
edges in \Fig{elepcomp}.

	The small lower limits for some of these distributions may
make it difficult to identify and measure the electrons in the
presence of the muon decay background. These limits are an
"accidental" consequence of the masses at this point; they decrease
slowly as $\sqrt s$ is increased.

\section{Error Analysis}

\begin{figure}[t]
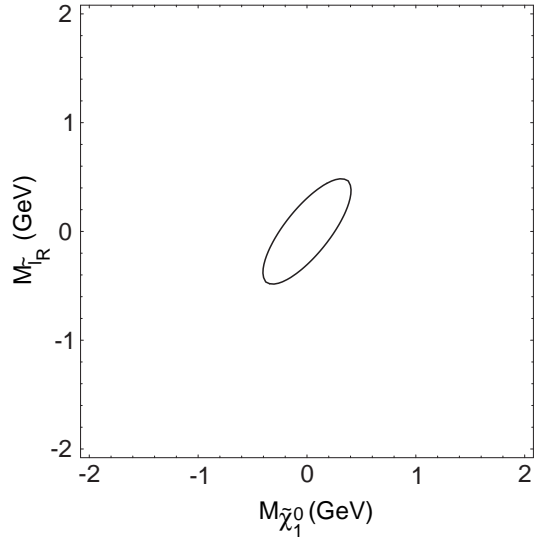

\dofig{2.75in}{errors1.ai}
\caption{Error ellipse in the $(M_\lsp, M_{\tilde\ell_R})$ plane from
measurements at the LHC and FMC at Point~5.\label{errors1}}
\end{figure}

	The easiest edge to detect in \Fig{eleperr} is the one at
$111.4\,\GeV$.  Statistically, one could detect this with bins of
$0.5\,\GeV$, ten times smaller, but detector resolution and possible
confusion from the two edges at about $100\,\GeV$ must be included. We
assume an error $\sigma_E = 1\,\GeV$. This edge is associated with the
$\ell_R$ sleptons, and its position is given by
$$
E_\ell^{\rm max} = 
{M_{\tilde\ell_R}^2 - M_\lsp^2 \over 4M_{\tilde\ell_R}^2}  
[\sqrt{s} + \sqrt{s - 4M_{\tilde\ell_R}^2}]
$$
The other $RR$ edge is at such low energy, $11.2\,\GeV$, that it will
probably be difficult to measure. The LHC can measure the endpoint of
the $\ell^+\ell^-$ mass spectrum and so determine
$$
M_{\ell\ell}^{\rm max} = 
M_{\tchi20} \sqrt{1 - {M_{\tilde\ell_R}^2 \over M_{\tchi20}^2}}
\sqrt{1 - {M_\lsp^2 \over M_{\tilde\ell_R}^2}}
$$
with an estimated error $\sigma_M = 1\,\GeV$ for $10\,\fbi$.\cite{HPSSY}
While one can estimate the masses independently, the results are not
so accurate, so we assume $M_{\tchi20} = 2 M_\lsp$. Then the $\chi^2$
error ellipse in the $(M_\lsp, M_{\tilde\ell_R}$ plane is given by
$$
\chi^2 = \sum_{ij} [
{1\over\sigma_E^2} {\partial E_{\ell}^{\rm max} \over
\partial M_i \partial M_j}
{1\over\sigma_M^2} {\partial M_{\ell\ell}^{\rm max} \over
\partial M_i \partial M_j}] \Delta M_i \Delta M_j
$$
The resulting error matrix is shown in \Fig{errors1} and represents a
significant improvement over the LHC alone.

	We next attempt to determine the $\tilde\ell_L$ mass by
measuring the edge at $178.3\,\GeV$ in \Fig{eleperr} from
$\tilde\ell_L^+\tilde\ell_L^-$ production. The error is probably about
one bin width, $5\,\GeV$. Unfortunately, the position of this edge,
$$ 
E_{\ell}^{\rm max} = {M_{\tilde\ell_L}^2 - M_\lsp^2 \over 4
M_{\tilde\ell_L}^2} [\sqrt{s} + \sqrt{s-4 M_{\tilde\ell_L}^2}] 
$$
turns out to be very insensitive to $M_{\tilde\ell_L}$ for these
values of the parameters; the derivative of the
$\tilde\ell_L\tilde\ell_L$ endpoint has a zero as a function of $\sqrt
s$ that happens to occur very close to $600\,\GeV$ for these masses.
Numerically,
$$
{d E_{\ell}^{\rm max} \over M_{\tilde\ell_L}} \approx 0.036\,.
$$
As a result the sensitivity is accidentally very poor. For "typical"
values the error on $M_{\tilde\ell_L}$ would be about three times that
on the edge. 

\section{Sleptons at LHC Point 3}

	The second LHC point at which the FMC could contribute is LHC
Point~3, a SUGRA point with $m_0=200\,\GeV$, $\mhalf=100\,\GeV$,
$A_0=0$, $\tan\beta=2$, and $\sgn\mu=-1$. These parameters were chosen
so that every accelerator could find something; in particular, LEP
recently announced discovery of the light Higgs at $68\,\GeV$. The LHC
can make many precise measurements at this point. However, since
sleptons do not occur in the cascade decays of gluinos and squarks,
they are not directly constrained.

\begin{figure}
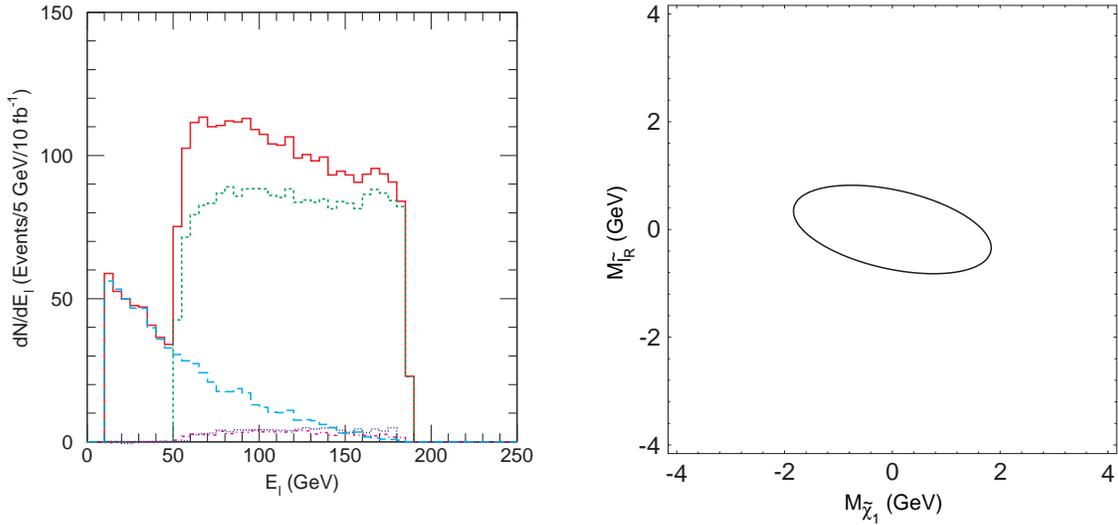

\hbox to \textwidth{\epsfxsize=3in\epsfbox{c3elepcomp.ai}\hfil
\epsfxsize=2.7in\epsfbox{errors4.ai}}
\caption{Composition of $E_\ell$ distribution for $\mu^+\mu^- + e^+e^-
- e^+\mu^- - e^-\mu^+$ at Point~3. The dashed curve is from
$\tilde\ell_R\tilde\ell_R$; the dotted curve is from
$\tilde\ell_L\tilde\ell_L$; the dashed-dotted curve is from
$\tilde\ell_R\tilde\ell_L$; and the long-dashed curve is from all
events with a $\tchi20$. The error ellipse is also
shown.\label{c3elepcomp}} 
\end{figure}

	\Fig{c3elepcomp} shows the lepton energy distribution for
Point~3 and its sources; compare with \Fig{elepcomp} for Point~5. The
sleptons are nearly degenerate at this point: $M_{\tilde\ell_R} =
206.5\,\GeV$ and $M_{\tilde\ell_R} = 215.7\,\GeV$. Hence all the
slepton edges are nearly degenerate; the $\tilde\ell_R\tilde\ell_R$
contribution dominates because the branching ratio for $\tilde\ell_R
\to \lsp \ell$ is nearly 100\%. There is also a contribution to
like-flavor dileptons from $\tchi20$ at this point, the long-dashed
curve in \Fig{c3elepcomp}. While the various contributions in probably
cannot be resolved, both the upper and the lower edges should be
measurable, allowing one to determine $M_{\tilde\ell}$ and $M_\lsp$
without additional assumptions.\cite{TFMYO}\cite{NFT} The resulting
error ellipse is shown assuming a measurement error of $1\,\GeV$ on
each edge. Clearly this case is much more favorable than that for
Point~5, although in part the difference is due to the fact that there
is less information on sleptons from the LHC.

	While polarization is not essential to detect the signal, it
would help to interpret it. In particular, the dominance of the
$\tilde\ell_R\tilde\ell_R$ contribution seen in \Fig{c3elepcomp} is
due to $\lsp$ exchange in the $t$-channel. Even rather modest beam
polarization would show that this contribution was dominant and so
provide another test of the SUSY model. The degree of polarization
required needs study but is probably much smaller than that needed to
suppress Standard Model backgrounds.

\bigskip

	This work was supported in part by the United States
Department of Energy under Contract DE-AC02-76CH00016.

\end{document}